# Resisting Selfish Mining Attacks in the Bicomp

Rui Tian[1,2] and Wei Gong[1,3]

[1] Chaincomp Technologies, Shenzhen 518000, China
[2] Beijing Engineering Research Center for IoT Software and Systems, Beijing University of Technology, Beijing 100024, China
[3] Research Center of Ubiquitous Sensor Networks, University of Chinese Academy of Sciences, Beijing 100049, China

## ABSTRACT

**Selfish mining, which is an attack on the integrity of the Bitcoin network, was first proposed by Cornell researchers Emin Gün Sirer and Ittay Eyal in 2013. Selfish mining attack also exists in most Nakamoto consensus protocols. Generally speaking, selfish mining strategy can comprise a Nakamoto consensus system with less than 25% mining power of the whole system. We have discussed how the Bicomp can resist selfish mining in our former paper "Bicomp: A Bilayer Scalable Nakamoto Consensus Protocol". In this technical report, we give a detailed derivation on the conditions a selfish attacker should meet to earn more revenues through selfish mining. And we also get a conclusion that through adjusting macroblock difficulties together with tenure lengths, the Bicomp protocol has high resistant towards selfish mining.**

## INDEX TERMS

Selfish mining, Nakamoto consensus, Proof-of-Work.

## 1. PRELIMINARIES

The Bicomp is a bilayer scalable Nakamoto consensus protocol we proposed in our previous work [1]. In Bicomp, two kinds of blocks are generated; that is, microblocks and macroblocks. The microblocks are used to concurrently package transactions in network, while macroblocks are used to package microblocks. Multicple macroblocks are connected consequently to form a trunk chain.

In Bicomp, as described in the security Analysis section (see section 4), selfish mining is largely restricted by tenure mechanisms and the rule of counting transaction diversities while selecting legitimate branches. However, that doesn't mean selfish mining does not work in our system. In this article, we will conduct an in-depth analysis on selfish mining in the Bicomp; that is, when, or under what condition a selfish miner can earn more revenue than mining honestly.

For simplicity, without loss of generality, we assume that, selfish attackers always account for a minority of nodes in the system; even they can possess relative higher computing power. Under this assumption, since selfish miners can only package transactions generated by themselves into the block, it is usually more difficult for them to generate more diverse transactions than honest nodes. This means that, selfish miners need always mine blocks at a larger height to guarantee the private branch will potentially replace the public branch.

x

The possible selfish mining strategy can be described as follow. We assume that nodes in the network are rational; that is, they try to maximize their revenue all the time. Suppose a selfish mining pool [2], called A, has total mining power of $\alpha$, while the others, called B, has mining power of $1-\alpha$. In round $n$, A successfully finds a macroblock header, called sMAB_header($n$), based on last macroblock, called MAB($n-1$), and distributes it to the network. After collects enough microblocks from the network, A detains the constructed macroblock, called sMAB($n$), and then mine subsequent header on sMAB ($n$) privately. At the same time, B never receives sMAB($n$) after receives sMAB_header($n$), so will reinitiate a leader-election process on MAB($n-1$) when time out. For simplicity, we assume that spreading time of macoblock header in the network is constant. So according to the formula (1) in [1], the timeout length that B waits for a macroblock equals leader's tenure length of each round. As a result, miner B again starts leader election process based on MAB($n-1$), while at the meantime selfish miner A mines on sMAB ($n$). During this process, A wins a leadership of one block height over B. Specifically, in round $n+1$, A mines blocks privately at height of $n+1$, while the honest node B mines at height of $n$. In following rounds, A can choose to reveal private branch and replace the public branch in case it found macroblock header for that round. Oppositely, if A failed to find macroblock header in any round, under the assumption that public macroblocks contains more diverse transactions than private ones, it no longer has chance to let the private branch replace the public one, and so will loss the revenue of finding the very beginning macroblock on the private branch, say sMAB($n$). So a miner will choose to mine privately if it will potentially win higher revenue than mine honestly.

In Bicomp, leaders earn revenues including mining rewards and transaction fees. For simplicity, we assume revenues are fixed, denote as $R$; even though honest miners are likely to earn more transaction fees than selfish miners through including diverse transactions in their macroblocks.

## 2. ANALYSIS AND DERIVATION

In this section, we will analyze the expected rewards for a selfish miner (or pool). We define the aforementioned system evolutions as state transitions, and construct state machine as shown in Figure 1. We define system state as the lead of private chain like in [2], but with different state transition conditions. State 0 implies that both A and B are mining on the same precedent macroblock. We don't split zero lead state further like in [2], since we assume a selfish attacker must ensure their advantage on the block height to earn extra revenues. State $n$ ($n>0$) is the state that private chain has a lead of $n$ in block height (including the newly found macroblock header) over the public branch. For instance, selfish miner A mines on height of $k+n$ blocks while the honest miner B mines on height of $k$.

**2.1 State Machine**

Mining events trigger state transitions in the state machine. When at state 0, if A finds next macroblock header while b doesn't, the lead of selfish branch increases by one, and the system becomes state 1. In case that B finds macroblock header in this round also, according our consumption, A will no longer gain block height advance so it will normally send constructed macroblock to public. When at state other than zero, whenever A finds next macroblock header in condition that B dosen't during a round, the lead of selfish branch increases by one. If A and B both find next macroblock header on their respective branches in the same round, the machine status remains unchanged. If B finds a block while A doesn't, the lead of selfish branch decreases by one.

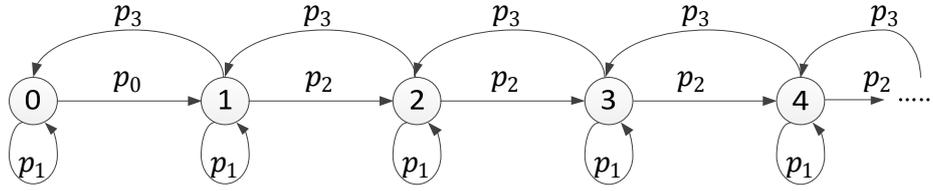

Fig.1 State machine with transition probabilities

In Bicomp, leader election must suspend during the tenure time of each round. Selfish miner can't privately keep mining all the time either, since nodes on the chain will check this timestamp before appending macroblocks. So either A or B experiences similar mining process, just mining events take place asynchronously. Before deriving state probabilities of the state machine, we define transition probabilities as below.

Table 1 Definitions of transition probabilities

| Denotation | Description |
| --- | --- |
| $p_0$ | The probability that A and B mine on the same branch, and A finds macroblock header. |
| $p_1$ | The probability that A and B both find macroblocks on their respective branch in one round |
| $p_2$ | The probability that A finds macroblock header on its private branch, while B fails in finding next macroblock header. |
| $p_3$ | The probability that A fails in finding next macroblock header on its private branch, while B finds next macroblock header. |

**2.2 State Probabilities and Revenue**

We denote state probability of state n as $q_n$. Then from the state machine, we can get following equations:

$$\begin{cases} p_0 q_0 = p_3 q_1 \\ \forall\, k > 0,\, p_2 q_k = p_3 q_{k+1} \\ \sum_{k=0}^{\infty} q_k = 1 \end{cases} \tag{1}$$

From equation (1), we can get:

$$q_0 = \frac{1}{p_0} \frac{p_0(p_3 - p_2)}{p_3 - p_2 + p_0} \tag{2}$$

$$q_1 = \frac{1}{p_3} \frac{p_0(p_3 - p_2)}{p_3 - p_2 + p_0} \tag{3}$$

......

$$\forall k > 1, q_k = \left(\frac{p_2}{p_3}\right)^{k-1} q_1 \tag{4}$$

The revenue can be obtained by selfish miner is intuitively in this state machine. Same as in [2], we can calculate revenues on events as bellow:

(1) When in state 1, if B finds a macroblock header, then A and B mines on the same height. A publishes the single hidden macroblock, and the macroblock with more non-overlapped transactions will be chosen to be appended on the chain, then the packer of the selected macroblock will win a revenue of R. We assume, without of loss of generality, the public and private macroblock have the same probabilities to be chosen, so in this case, A and B both win an expected revenue of 0.5R.

(2) When in state 2, in case that B finds a new macroblock header, A publishes the previous detained macroblock and the newly found macroblock header simultaneously. Since the private branch are longer than the public branch, the public branch will be discarded and then A wins a revenue of two.

(3) When in states those are larger than 2, and in case that B finds a new block header, the gap between the secret and public branches reduces by 1. A can choose to publish its most precedent private macroblocks or macroblock header, or do nothing. Since A has longer branch hidden and will eventually win in the fork competition, we suppose A wins a revenue of R in this case, although it can not get the revenue right away.

(4) When at any state and A finds next macroblock header, the lead of private branch has on the public branch increases by 1. The revenue with this macroblock header will de determined later.

From above analysis, we can calculate the revenue of A and B as:

$$r_a = \left(q_1 p_3 * 0.5 + q_2 p_3 * 2 + \sum_{k=3}^{\infty} q_k p_3 * 1\right) * R \tag{5}$$

$$r_b = q_1 p_3 * 0.5 * R \tag{6}$$

The revenue rate ratio of the selfish miner is as:

$$R_a = \frac{r_a}{r_a + r_b} = \frac{2 - 2q_0 - q_1 + 2q_2}{2 - 2q_0 + 2q_2} \tag{7}$$

Since A has mining power of α, if it mines honestly, it can obtain revenue rate of α. So the way to prompt selfish mining is to let the $R_a$ less than α. that is:

$$\frac{2 - 2q_0 - q_1 + 2q_2}{2 - 2q_0 + 2q_2} < \alpha \tag{8}$$

From equations (2)~(4), and (8), we can observe that, in the Bicomp, by varying $p_0 \sim p_3$, the revenue rate ratio of the selfish pool can be controlled. The probabilities $p_0 \sim p_3$ correspond to mining activities listed in Table 1. According to the mining rules in Bicomp, the tenure length and macroblock POW difficulty, which are two configurable parameters of the system, together decide these probabilities. It implies that, in condition that one of the two parameters is determined, we can adjust the other parameter to make the system resist to specific selfish mining attack requirement. That is, the minimum mining power required to make a selfish miner earn extra revenue. When implemented, the inequality (8) can be solved through Monte Carlo method, and we will not make further derivation in this article.

## 3. FIXING THE BICOMP

From above discussions, we know that the system's resistance towards selfish mining can be configured through adjusting tenure length and macroblock difficulty. However, the range of values of these two parameters has been reduced a lot to this end. We propose a fixing mechanism to the Bicomp in this section, to make the system has stronger resistance to selfish mining, without substantially modifying the operation mechanism of the system.

In section I, we propose that, when the selfish pool A publishes a macroblock header first and then detains the constructed macroblock, the honest miners B will reinitiate a leader electing process due to timeout. By doing so, selfish pool can easily get a lead of 1 block in block height than honest miners. Considering in the Bicomp, multiple leaders are allowed in each round, we intuitively propose that, generating more headers in each round, and let miners to mine on any existing headers. So even if selfish miners detain constructed macroblock, honest miners can also mine on other optional macroblocks in next round and thus have the same mining heights as selfish miners. According to the Bicomp, the branches with most non-overlapped transactions will be chosen as the trunk chain, so the secret branch has little chance to be chosen as trunk chain since less non-overlapped transactions are included in the branch, as we assumed in section I.

## 4. CONCLUSION

In this article, we have analyzed in detail the selfish mining resistance of the Bicomp protocol. Through modeling the system as a state machine, and analyze different mining activities that lead state transition, we come to the conclusion that the Bicomp can adjust its resistance towards selfish mining attack by varying macroblock POW difficulty and tenure length parameters. We also present a modification to the Bicomp protocol without substantially modifying the operation mechanism of the system.